\documentclass[10pt,a4paper,final]{iopart}

\usepackage{iopams}  
\usepackage{graphicx,colordvi}
\usepackage{amssymb,slashed,cite}
\expandafter\let\csname equation*\endcsname\relax
\expandafter\let\csname endequation*\endcsname\relax
\usepackage{amsmath}

\usepackage{booktabs,tabulary}
\usepackage{dsfont}
\usepackage{color}
\usepackage[breaklinks=true,colorlinks=true,linkcolor=blue,urlcolor=blue,citecolor=blue]{hyperref}

\newcommand{\diff}{\text{d}}
\newcommand{\MeV}{\,\text{MeV}}
\newcommand{\GeV}{\,\text{GeV}}
\renewcommand{\Im}{\text{Im}\,}

\newcommand{\mpi}{M_\pi} 

\newcommand{\mN}{m_N}

\newcommand{\Order}{\mathcal{O}}

\newcommand{\zzeta}{\boldsymbol{\zeta}}
\newcommand{\beq}{\begin{equation}}
\newcommand{\eeq}{\end{equation}}

\newcommand{\toright}[1]{\hspace*{\fill}{\footnotesize{#1}}}

\begin{document}

\title[Extracting the $\sigma$-term from low-energy pion--nucleon scattering]{\toright{\textnormal{INT-PUB-17-019}}\\[0.1cm]Extracting the $\boldsymbol{\sigma}$-term from low-energy pion--nucleon scattering}

\author{Jacobo Ruiz de Elvira$^{1}$, Martin Hoferichter$^2$, \\Bastian Kubis$^{3}$ and Ulf-G.\ Mei{\ss}ner$^{3,4}$}
\address{$^1$ Albert Einstein Center for Fundamental Physics, Institute for Theoretical Physics,
University of Bern, Sidlerstrasse 5, CH--3012 Bern, Switzerland}
\address{$^2$ Institute for Nuclear Theory, University of Washington, Seattle, WA 98195-1550, USA}
\address{$^3$ Helmholtz--Institut f\"ur Strahlen- und Kernphysik (Theorie) and
   Bethe Center for Theoretical Physics, Universit\"at Bonn, D--53115 Bonn, Germany}
\address{$^4$ Institut f\"ur Kernphysik, Institute for Advanced Simulation, and 
   J\"ulich Center for Hadron Physics, Forschungszentrum J\"ulich, D--52425  J\"ulich, Germany}

\eads{\mailto{elvira@itp.unibe.ch}, \mailto{mhofer@uw.edu}, \mailto{kubis@hiskp.uni-bonn.de} and \mailto{meissner@hiskp.uni-bonn.de}}

\begin{abstract}
We present an extraction of the pion--nucleon ($\pi N$) scattering lengths from low-energy $\pi N$ scattering, by fitting a representation based on Roy--Steiner equations to the
low-energy data base. We show that the resulting values confirm the scattering-length determination from pionic atoms, and discuss the stability of the fit results regarding electromagnetic corrections and experimental normalization uncertainties in detail.  
Our results provide further evidence for a large $\pi N$ $\sigma$-term, $\sigma_{\pi N}=58(5)\MeV$, in agreement with, albeit less precise than, the determination from pionic atoms. 
\end{abstract}

\pacs{13.75.Gx, 11.55.Fv} 
\vspace{2pc}
\noindent{\it Keywords}: Pion--baryon interactions, Dispersion relations, Pion--nucleon $\sigma$-term 

\section{Introduction}

The $\pi N$ $\sigma$-term, $\sigma_{\pi N}$, encodes crucial information about the scalar nucleon matrix elements of up- and down-quarks $\langle N|m_q\bar q q|N\rangle$ for $q=u,d$~\cite{Crivellin:2013ipa}.
These scalar couplings not only emerge as fundamental QCD parameters, measuring the amount of the nucleon mass generated by up- and down-quarks, but also determine nucleon matrix elements 
required in the search for physics beyond the Standard Model, e.g.\ in the direct detection of dark matter~\cite{Bottino:1999ei,Bottino:2001dj,Ellis:2008hf}, 
searches for lepton flavor violation in $\mu\to e$ conversion in nuclei~\cite{Cirigliano:2009bz,Crivellin:2014cta}, or electric dipole moments~\cite{Engel:2013lsa,deVries:2015gea,deVries:2016jox,Yamanaka:2017mef}.
Such nucleon matrix elements corresponding to an interaction channel not present in the Standard Model as an external current are difficult to constrain from experiment, often leaving 
ab-initio calculations in lattice QCD as the only viable approach. The scalar couplings of the lightest quarks, however, form an important exception: due to the $SU(2)$ Cheng--Dashen low-energy theorem~\cite{Cheng:1970mx,Brown:1971pn} there is a rigorous relation between $\sigma_{\pi N}$ and the $\pi N$ scattering amplitude that allows one to extract the $\sigma$-term from low-energy $\pi N$ phenomenology with high precision, providing a rare opportunity to test lattice-QCD calculations of nucleon matrix elements in experiment.

In practice, the low-energy theorem requires the analytic continuation of the $\pi N$ amplitude into the subthreshold region, which can be achieved in a stable way based on dispersion relations, given sufficient data input for low-energy $\pi N$ scattering. This analytic continuation was performed in~\cite{Gasser:1988jt,Gasser:1990ce,Gasser:1990ap} based on the Karlsruhe--Helsinki (KH80) partial-wave analysis~\cite{Koch:1980ay,Hoehler}, leading to $\sigma_{\pi N}\sim 45\MeV$~\cite{Gasser:1990ce}, while, within the same formalism, the more recent SAID/GWU partial-wave analysis~\cite{Pavan:2001wz} suggested a significantly larger value $\sigma_{\pi N}=64(8)\MeV$. Over the past years, the combination of partial-wave dispersion relations together with unitarity and crossing symmetry in the framework of Roy--Steiner (RS) equations~\cite{Ditsche:2012fv,Hoferichter:2012wf,Hoferichter:2015dsa,Hoferichter:2015tha,Hoferichter:2015hva,Hoferichter:2016ocj,Hoferichter:2016duk,Siemens:2016jwj} further
sharpened the relation between $\sigma_{\pi N}$ and low-energy $\pi N$ scattering. In particular, the strong constraint imposed by the RS equations on the energy dependence of the scattering amplitude allows one to eliminate the need for experimental input for a particular $P$-wave scattering volume~\cite{Gasser:1988jt,Gasser:1990ce} and instead derive a direct relation to the $S$-wave scattering lengths $a^{I_s}$, $I_s=1/2,3/2$,
\begin{align}
\label{sigma_piN_lin}
\sigma_{\pi N}&=59.1(3.1)\MeV + \sum_{I_s}c_{I_s}\big(a^{I_s}-\bar a^{I_s}\big),\notag\\
c_{1/2}&=0.242\MeV\times 10^3\mpi,\qquad c_{3/2}=0.874\MeV\times 10^3\mpi,
\end{align}
where the reference values $\bar a^{I_s}$ refer to the values extracted from high-precision data on pionic atoms taken at PSI~\cite{Strauch:2010vu,Hennebach:2014lsa,Gotta:2008zza},
leading to~\cite{Baru:2010xn,Baru:2011bw,Hoferichter:2015hva}
\begin{align}
\label{scatt_length_pionic}
\bar a^{1/2}&=169.8(2.0)\times 10^{-3}\mpi^{-1},\notag\\ 
\bar a^{3/2}&=-86.3(1.8)\times 10^{-3}\mpi^{-1}.
\end{align}
The increase of the resulting $\sigma$-term
\beq
\label{sigma_RS}
\sigma_{\pi N}=59.1(3.5)\MeV
\eeq
compared to~\cite{Gasser:1990ce} can be attributed in part to isospin-breaking corrections to the low-energy theorem~\cite{Hoferichter:2015dsa}, but mainly to improved experimental input. As demonstrated in detail in~\cite{Hoferichter:2015dsa,Hoferichter:2015hva}, using the KH80 scattering lengths~\cite{Hoehler}
\begin{align}
\label{KH80}
a^{1/2}_\text{KH80}&=173(3)\times 10^{-3}\mpi^{-1},\notag\\ 
a^{3/2}_\text{KH80}&=-101(4)\times 10^{-3}\mpi^{-1}
\end{align}
instead of~\eqref{scatt_length_pionic} indeed reproduces the small sigma term of~\cite{Gasser:1990ce} (note that the detailed comparison involves the consideration of isospin-breaking corrections, see also~\cite{Gasser:2002am,Hoferichter:2009ez,Hoferichter:2009gn,Hoferichter:2012bz}).
A similar conclusion has been reached within chiral perturbation theory (ChPT)~\cite{Fettes:2000xg,Alarcon:2011zs}: the extracted 
value of $\sigma_{\pi N}$ varies according to the partial-wave analysis used to determine the low-energy constants, but demanding that the scattering lengths agree with~\eqref{scatt_length_pionic}
prefers the SAID/GWU solution, with a result $\sigma_{\pi N}=59(7)\MeV$~\cite{Alarcon:2011zs} in perfect agreement with~\eqref{sigma_RS}. 

\begin{table}
\centering
\renewcommand{\arraystretch}{1.3}
\begin{tabular}{cccc}
\toprule
Collaboration & $\sigma_{\pi N}\, [\MeV]$ & Reference &\\\midrule
BMW & $38(3)(3)$ & \cite{Durr:2015dna} & $3.8\sigma$\\
$\chi$QCD & $45.9(7.4)(2.8)$ & \cite{Yang:2015uis} & $1.5\sigma$\\
ETMC & $37.2(2.6)\big(^{4.7}_{2.9}\big)$ & \cite{Abdel-Rehim:2016won} & $3.4\sigma$\\
RQCD & $35(6)$ & \cite{Bali:2016lvx} & $3.5\sigma$\\
\bottomrule
\end{tabular}
\caption{Recent lattice results for $\sigma_{\pi N}$. The tension with~\eqref{sigma_RS} is given in the last column (errors added in quadrature).}
\label{table:lattice}
\renewcommand{\arraystretch}{1.0}
\end{table}

While the phenomenological determination therefore appears internally consistent, the comparison to recent lattice-QCD calculations at or near physical quark masses~\cite{Durr:2015dna,Yang:2015uis,Abdel-Rehim:2016won,Bali:2016lvx} reveals a tension with significance as given in Table~\ref{table:lattice}.
A natural strategy to try and resolve this discrepancy involves the $\pi N$ scattering lengths~\eqref{scatt_length_pionic}: in case there were some overlooked systematic effect, this could result in a shift in the $\sigma$-term as well, so that it would be desirable to obtain an independent check of the pionic-atom result.
Such a test could be provided by a lattice calculation of the scattering lengths themselves, 
although a relatively high accuracy would be required to make the result conclusive~\cite{Hoferichter:2016ocj}. 
Alternatively, we pursue another data-driven approach in this paper, based on the comparison to the data base for low-energy $\pi N$ scattering: 
a remarkable feature of the RS representation of the $\pi N$ amplitude concerns the fact that
it is completely determined once the scattering lengths are fixed, so that the derivation of~\eqref{sigma_RS} does not  
make use of actual $\pi N$ scattering data except for the high-energy tails of the dispersive integrals. 
Therefore, the low-energy $\pi N$ cross section is efficiently parameterized in terms of the scattering lengths only,
which can then be extracted by fitting the corresponding representation to the data base. 

Apart from the use of a RS representation, our fit strategy differs from recent fits to $\pi N$ scattering data based on chiral amplitudes~\cite{Wendt:2014lja,Siemens:2016hdi,Siemens:2017opr} in two
major ways: first, we do not use the experimental normalizations determined by the SAID/GWU group~\cite{Workman:2012hx,SAID}, but extract these normalizations directly from the data base (these scale factors have also been reconsidered recently in~\cite{Matsinos:2016fcd}, suggesting that systematic uncertainties in the low-energy behavior of the SAID/GWU solution are underestimated). 
Second, the electromagnetic corrections from~\cite{Tromborg:1973zt,Tromborg:1975mn,Tromborg:1976bi,Tromborg:1977da} beyond the largely unambiguous Coulomb part are included in the error estimate, and we argue how our approach conservatively maps onto the isospin conventions for the scattering lengths. In this way, the two main sources of uncertainty when fitting to the $\pi N$ data base should be adequately addressed.

After reviewing the essential features of RS equations in Sect.~\ref{sec:RS} and electromagnetic corrections in Sect.~\ref{sec:EM}, Sect.~\ref{sec:fit} will be devoted to the details of the fit,
including the role of the uncertainties related to the experimental normalizations. 
We summarize the consequences for $\sigma_{\pi N}$ in Sect.~\ref{sec:sigma_term}, before concluding in Sect.~\ref{sec:con}.

\section{Roy--Steiner representation}
\label{sec:RS}

For a detailed account of $\pi N$ RS equations we refer to~\cite{Hoferichter:2015hva}, here we review the salient properties needed for the fit
to the $\pi N$ data base. The RS equations for the $s$-channel partial waves $f_{l\pm}^{I_s}$, with angular momentum $l\pm 1/2$, take the form
\begin{align}
\label{sRSpwhdr}
f^{I_s}_{lI}(W)&=N^{I_s}_{lI}(W)
+\frac{1}{\pi}\int\limits^{\infty}_{W_+}\diff W'\sum_{l'I_s'I'}K_{ll'II'}^{I_sI'_s}(W,W')\,\Im f^{I'_s}_{l'I'}(W')\notag\\
&+\frac{1}{\pi}\int\limits^{\infty}_{4\mpi^2}\diff t'\sum_{JI'}
G^{I_s}_{lJII'}(W,t')\,\Im f^J_{I'}(t'),
\end{align}
where $I,I'=\pm$, $\mN$, $\mpi$ denote nucleon and pion masses, $W=\sqrt{s}$, $s$, $t$ are Mandelstam variables, $W_+=\mN+\mpi$, $N^{I_s}_{l\pm}$ denotes the partial-wave projection of the Born terms, $f^J_\pm(t)$ refers to the partial waves for the crossed-channel process $\pi\pi\to \bar NN$ (with total angular momentum $J$ and parallel/anti-parallel nucleon--antinucleon helicities), and $K_{ll'II'}^{I_sI'_s}$, $G^{I_s}_{lJII'}$ are analytically known kernel functions.
The $s$-channel kernels include a diagonal term
\beq
K_{ll'II'}^{I_sI'_s}(W,W')=\frac{\delta_{ll'}\delta_{I_sI_s'}\delta_{II'}}{W'-W}+\ldots,
\eeq
which together with the unitarity relation
\beq
\Im f^{I_s}_{l\pm}(W)=q\big|f^{I_s}_{l\pm}(W)\big|^2, \qquad q=\frac{\sqrt{\lambda\big(s,\mN^2,\mpi^2\big)}}{2\sqrt{s}}
\eeq
being the center-of-mass momentum [with the K\"all\'en function $\lambda(a,b,c)=a^2+b^2+c^2-2(ab+ac+bc)$],
and its elastic solution
\beq
f^{I_s}_{l\pm}(W)=\frac{e^{i\delta^{I_s}_{l\pm}(W)}}{q}\sin \delta^{I_s}_{l\pm}(W),
\eeq
implies a set of non-linear integral equations for the phase shifts $\delta^{I_s}_{l\pm}$. 

Following the conventions of~\cite{Hoehler}, the differential cross section $\diff\sigma/\diff\Omega$ and polarization $P$ can be expressed in terms of spin-flip ($H$) and non-flip ($G$) amplitudes according to
\beq
 \frac{\diff\sigma}{\diff\Omega}=\Big(|G|^2+|H|^2\Big)\frac{q_\text{f}}{q_\text{i}},\qquad
 P=\frac{2\Im\big(G H^*\big)}{|G|^2+|H|^2},
\eeq
where $q_\text{f/i}$ refer to final-/initial-state center-of-mass momenta, which differ in the case of the charge-exchange channel. Unless otherwise noted, all kinematic quantities will refer to their isospin-limit values defined by the charged-particle masses. 
The amplitudes are related to the partial waves by means of
\begin{align}
\label{GH_PW}
G(s,t)&=\sum\limits_{l=0}^\infty \big[(l+1)f_{l+}(W)+l f_{l-}(W)\big]P_l(\cos\theta),\notag\\
H(s,t)&=\sum\limits_{l=1}^\infty \big[f_{l+}(W)-f_{l-}(W)\big]P'_l(\cos\theta)\sin\theta,
\end{align}
with scattering angle $z=\cos\theta$, $t=-2q^2(1-z)$, and the standard Legendre polynomials $P_l$ as well as their derivatives $P'_l$.

The polarization observables $P$ do not receive any contributions from $S$-waves and hence their dependence on the scattering lengths is expected to be minor. 
For this reason we will restrict ourselves to study differential cross section data. 
As shown in~\cite{Hoferichter:2015hva}, the corresponding constraint on the energy dependence of $f^{I_s}_{l\pm}$ fixes the low-energy behavior of the $G$ and $H$ amplitudes completely once the $S$-wave scattering lengths are specified, 
which are related by the normalization $f^{I_s}_{0+}(W_+)=a^{I_s}$. In close analogy to~\eqref{sigma_piN_lin}, the cross section solution can thus be written in a linearized form
\beq\label{eq:dcross}
\frac{\diff\sigma}{\diff\Omega}(W,z)=\frac{\diff\bar{\sigma}}{\diff\Omega}(W,z) +\sum_{I_s}\big(a^{I_s}-\bar a^{I_s}\big)d_{I_s}(W,z),
\eeq
where the energy- and angle-dependent corrections $d_{I_s}$ take the role of the $c_{I_s}$ coefficients. 
Such a representation fit to data then determines $a^{I_s}$, while in~\cite{Hoferichter:2015hva} the scattering lengths were constrained to the pionic-atom values $\bar a^{I_s}$ and considered as external input. 
The difference between these strategies can be illustrated using the example of $\pi\pi$ scattering, where again the scattering lengths are the essential free parameters~\cite{Ananthanarayan:2000ht}.
To determine these parameters, one can either provide additional external input, in the form of two-loop ChPT~\cite{Colangelo:2001df}, or low-energy data, 
in this case most prominently from $K_{l4}$ decays~\cite{Colangelo:2001sp,GarciaMartin:2011cn}, leading to two independent determinations of the $\pi\pi$ scattering lengths that permit a powerful consistency check.
The aim of the present paper is to establish a second determination of the $\pi N$ scattering lengths, in analogy to the $K_{l4}$ fit~\cite{Colangelo:2001sp,GarciaMartin:2011cn}, 
in order to obtain a similar consistency check for the $\pi N$ case as well.

In practice, the fitting procedure is less straightforward than described so far. First, the  RS equations~\eqref{sRSpwhdr} are only valid in a finite energy region below $W_\text{m}=1.38\GeV$, and only a finite number of partial waves can be retained in the solution. This implies that for energies above $W_\text{m}$ as well as higher partial waves input from existing partial-wave analyses has to be used, which introduces uncertainties. Similarly, the solution of the analogous set of integral equations for the $t$-channel problem involves uncertainties that propagate towards the $s$-channel solution.
While at low energies the scattering lengths indeed constitute the only free parameters, the remaining uncertainties in the RS solution become increasingly important for higher energies. For this reason, we restrict the analysis to low-energy data for which the sensitivity to the scattering lengths is greatest, ignoring data once the uncertainties from other sources become dominant.

Finally, although the RS equations~\eqref{sRSpwhdr} in principle couple different isospin amplitudes, at low energies the sensitivity to the main isospin component is by far most pronounced, i.e.\ for the experimentally accessible channels
\begin{align}
 \pi^+p\to\pi^+ p&: \quad a^{3/2},\notag\\
 \pi^-p\to\pi^- p&: \quad \frac{1}{3}\Big(2a^{1/2}+a^{3/2}\Big),\notag\\
 \pi^-p\to\pi^0 n&: \quad -\frac{\sqrt{2}}{3}\Big(a^{1/2}-a^{3/2}\Big).
\end{align}
Since the discrepancy between the pionic-atom scattering lengths~\eqref{scatt_length_pionic} and the KH80 ones~\eqref{KH80} resides almost exclusively in the $I_s=3/2$ channel, we thus expect the $\pi^+p\to\pi^+p$ reaction to be most powerful in discriminating between the two sets.  

\section{Electromagnetic corrections}
\label{sec:EM}

At low energies and in forward direction, $\pi N$ cross sections are strongly affected by electromagnetic interactions.
The set of electromagnetic corrections widely used in $\pi N$ scattering dates back to~\cite{Tromborg:1973zt,Tromborg:1975mn,Tromborg:1976bi,Tromborg:1977da}, culminating in the prescription how to apply these corrections in actual data analyses as summarized in~\cite{Tromborg:1976bi}. We first review these corrections, before detailing the consequences for our fit.

The dominant radiative corrections arise from single-photon-exchange diagrams, which for $\pi^\pm p\to\pi^\pm p$ produce the additional Coulomb contributions
\begin{align}
\label{Coulomb}
 G_\text{C}(s,t)&=\pm\bigg\{\bigg(\frac{2q\gamma}{t}+\frac{\alpha}{2W}\frac{W+\mN}{E+\mN}\bigg)F_1(t)\notag\\
 &\qquad+\bigg(W-\mN+\frac{t}{4(E+\mN)}\bigg)\frac{\alpha}{2\mN W}F_2(t)\Bigg\}F_\pi^V(t)e^{\pm i\phi_\text{C}(s,t)},\notag\\
 H_\text{C}(s,t)&=\pm\frac{\alpha F_\pi^V(t)}{2W\tan\frac{\theta}{2}}\Bigg\{\frac{W+\mN}{E+\mN}F_1(t)
 +\frac{1}{\mN}\bigg(W+\frac{t}{4(E+\mN)}\bigg)F_2(t)\bigg\},
\end{align}
to be added to~\eqref{GH_PW}. Here, $F_\pi^V$ denotes the electromagnetic form factor of the pion, $F_{1/2}$ the Dirac/Pauli form factor of the proton, $e^2=4\pi\alpha$,
\beq
E=\frac{s+\mN^2-\mpi^2}{2W},\qquad \gamma=\alpha \frac{s-\mN^2-\mpi^2}{2q W},
\eeq
and $\phi_\text{C}$ ensures that the most singular pieces are correct at $\Order(\alpha^2)$~\cite{Tromborg:1973zt}
\beq
\phi_\text{C}(s,t)=-\gamma\log\sin^2\frac{\theta}{2}+\gamma\int\limits_{-4q^2}^0\frac{\diff t'}{t'}\Big(1-F_1(t')F_\pi^V(t')\Big)-2\gamma C_E,
\eeq
with the Euler--Mascheroni constant $C_E$. 
We emphasize that the form factor parameterizations (of dipole form) employed in~\cite{Tromborg:1973zt,Tromborg:1975mn,Tromborg:1976bi} are not very accurate compared to today's standards. However, using modern input we checked that for the data of interest in the present study, they are probed at such small values of $t$ that improved form factors do not affect the extracted strong amplitudes in any noticeable way. 

In addition to the direct contribution to observables, the Coulomb amplitudes affect the partial waves by means of photon exchange in the initial/final state, which produces the Coulomb phase shifts 
\beq
 f_{l\pm}\to f_{l\pm}e^{iQ\Sigma_{l\pm}},
\eeq
where $Q$ refers to the sum of the products of particle charges in the initial and final states,
\beq
Q_{\pi^\pm p\to\pi^\pm p}=\pm 2,\qquad Q_{\pi^-p\to\pi^0 n}=-1, 
\eeq
and the phase shifts follow from the (regularized) partial-wave projection of the Coulomb amplitudes, i.e.\ at $\Order(\alpha)$~\cite{Tromborg:1973zt}
\begin{align}
\label{Coulomb_phase}
 \Sigma_{l\pm}&=-\gamma C_E+\frac{q}{2}\int\limits_{-1}^1\diff z F_\pi^V(t)\bigg(\frac{2q\gamma}{t}F_1(t)\big(P_l(z)-1\big)-\frac{\alpha}{2W}F_2(t)P_l(z)\bigg)\notag\\
 &\pm \frac{\alpha q}{2W(2l+1\pm1)}
 \int\limits_{-1}^1\diff zF_\pi^V(t)\big(P'_l(z)+P'_{l\pm 1}(z)\big)\notag\\
 &\qquad\times \bigg\{\frac{W+\mN}{E+\mN}F_1(t)+\frac{1}{\mN}\bigg(W+\frac{t}{4(E+\mN)}\bigg)F_2(t)\bigg\}.
\end{align}

Beyond these Coulomb effects, the corrections parameterized in~\cite{Tromborg:1976bi} at the level of hadronic phase shifts and inelasticity parameters 
include further radiative corrections from virtual photons and bremsstrahlung, from mass differences in nucleon Born terms, from different $\pi N$ coupling constants, and from inelasticities related to the $n\gamma$ channel, while contributions from short-range photons cannot be addressed within this framework~\cite{Tromborg:1977da}.
For the present paper, it is crucial to adopt isospin conventions that match onto~\cite{Hoferichter:2015dsa,Hoferichter:2015hva}, to ensure consistency with the isospin-breaking corrections in the low-energy theorem. 
Therefore, the isospin limit should be defined by the elastic channels, and all effects from virtual photons removed from the fit scattering lengths.
To a large extent the corrections from~\cite{Tromborg:1976bi} are consistent with this set-up, e.g.\ the effect from hard photons parameterized in terms of the low-energy constants $f_1$, $f_2$ was not removed from the scattering lengths in~\cite{Hoferichter:2015dsa,Hoferichter:2015hva}. On the other hand, only the infrared properties of virtual-photon loops are unambiguously calculable in QED, so that 
virtual-photon effects cannot be captured fully consistently with ChPT, as illustrated by the need to introduce counterterms to cure the ultraviolet divergences.

For these reasons we will adopt the following strategy: central values will be quoted for the full corrections from~\cite{Tromborg:1976bi} and the fit results will be interpreted as virtual-photon-subtracted scattering lengths in the sense of~\cite{Hoferichter:2015dsa,Hoferichter:2015hva}, but the difference to the fit results obtained by switching off all corrections except for~\eqref{Coulomb} and~\eqref{Coulomb_phase} will be quoted as an additional systematic uncertainty, as will be the shifts expected from ChPT in the scattering lengths when subtracting virtual-photon effects. 
The latter amount to
\begin{align}
\label{ChPT_virtual}
\Delta a_{\pi^+p\to\pi^+p}^\gamma&=-1.3(0.9)\times 10^{-3}\mpi^{-1},\notag\\
\Delta a_{\pi^-p\to\pi^-p}^\gamma&=0.8(0.9)\times 10^{-3}\mpi^{-1},\notag\\
\Delta a_{\pi^-p\to\pi^0n}^\gamma&=-0.3(0.4)\times 10^{-3}\mpi^{-1}.
\end{align}

\section{Fit to the pion--nucleon data base}
\label{sec:fit}

We use the data base compiled by the GW group~\cite{Workman:2012hx,SAID} as a starting point for our fit, with electromagnetic corrections as detailed in the previous section. However, the main experimental uncertainty beyond the statistical errors quoted by each experiment concerns the normalizations, which in~\cite{Workman:2012hx,SAID} have been determined in the context of their $\pi N$ partial-wave solution. To avoid any bias, we do not use these GW normalizations, but instead consider them as additional fit parameters (besides the scattering lengths).
Further bias related to normalization uncertainties might arise in a $\chi^2$ minimization, see~\cite{DAgostini:1993arp}, but this can be avoided based on the following iterative strategy~\cite{Ball:2009qv}.

For $k=1,\ldots,N$ experiments, with $N_k$ data points, we are given sets of data $\{W_i^k,\sigma_i^k,\Delta\sigma_i^k\}$, with center-of-mass energies $W_i^k$, $i=1,\ldots,N_k$, and for simplicity we denote all observables by $\sigma$, whose statistical errors $\Delta\sigma$ are assumed to be uncorrelated (in practice, the differential cross sections also depend on angles $z_i^k$). Suppose we have some initial guess for the scattering length $a_0$ as well as for the normalizations $\zeta_{0,k}$ and their uncertainties $\Delta\zeta_{0,k}$ (which we will collect in vectors $\zzeta_0$ and $\Delta\zzeta_0$), then an unbiased solution can be obtained by minimizing
\begin{align}
\label{fit_strategy}
 \chi^2(a,a_0,\zzeta,\zzeta_0,\Delta\zzeta_0)&=\sum_{k=1}^N\chi^2_k(a,a_0,\zzeta,\zzeta_0,\Delta\zzeta_0),\notag\\
 \chi^2_k(a,a_0,\zzeta,\zzeta_0,\Delta\zzeta_0)&=\notag\\
 &\hspace{-20pt}\sum_{i,j=1}^{N_k}\Big(\zeta_k^{-1}\sigma(W_i^k,a)-\sigma_i^k\Big)\big(C_k^{-1}(a_0,\zzeta_0,\Delta\zzeta_0)\big)_{ij}\Big(\zeta_k^{-1}\sigma(W_j^k,a)-\sigma_j^k\Big),\notag\\
 \big(C_k(a_0,\zzeta_0,\Delta\zzeta_0)\big)_{ij}&=\delta_{ij}\big(\Delta\sigma_i^k\big)^2+\zeta_{0,k}^{-1}\sigma(W_i^k,a_0)\zeta_{0,k}^{-1}\sigma(W_j^k,a_0)\bigg(\frac{\Delta\zeta_{0,k}}{\zeta_{0,k}}\bigg)^2\notag\\
 &=\delta_{ij}\big(\Delta\sigma_i^k\big)^2+\sigma(W_i^k,a_0)\sigma(W_j^k,a_0)\bigg(\frac{\Delta\zeta_{0,k}}{\zeta_{0,k}^2}\bigg)^2,
\end{align}
with respect to the scattering length $a$ and the normalizations $\zeta_k$. At the minimum, the new errors $\Delta\zzeta$ are then derived from the inverse of the Hessian, and the system can be iterated until convergence. 
By definition, the different $\zeta_k$ will be uncorrelated, and we checked that the correlations between scattering length and normalizations are negligibly small.

As described in Sect.~\ref{sec:RS}, we restrict the analysis to low energies where the sensitivity to the scattering lengths in the RS solution exceeds the size of the uncertainties from other sources.
In more detail, the maximum energy for a given channel $I_s$ is taken as the maximum energy $W_\text{max}$ for which the relation 
\begin{equation}
  \left\vert \frac{\diff\sigma}{\diff\Omega}(W,z,\bar a)-\frac{\diff\sigma}{\diff\Omega}(W,z,a_\text{KH80})\right\vert\leq\Delta\bigg(\frac{\diff\sigma}{\diff\Omega}\bigg)(W,z,\bar a)
\end{equation}
holds for every scattering angle $z$, with scattering lengths as defined in~\eqref{scatt_length_pionic} and~\eqref{KH80}.  
$\Delta(\diff\sigma/\diff\Omega)$ denotes the remaining RS uncertainties in the cross section central solution, as detailed in~\cite{Hoferichter:2015hva}. 
For instance, it includes the effect of the uncertainty in the $\pi N$ coupling constant, from flat-minima directions in the fit parameter space of the RS solution, 
the variation of the matching conditions, $t$-channel input, and $s$-channel $l\geq 2$ partial waves. 
We also studied the effect of the RS theory uncertainties in the fit by comparing the fit results for different solutions from~\cite{Hoferichter:2015hva}, but the resulting differences
proved negligible.

For the $\pi^+p\to\pi^+p$ channel, we find $T_\pi^\text{max}=51.5\MeV$, with 
\beq
T_\pi=\frac{W^2-(\mN+\mpi)^2}{2 \mN}
\eeq
the  kinetic energy of the incoming pion in the lab frame.
This energy range accounts for 32 experimental measurements and 464 data points.  
In the $\pi^-p\to\pi^-p$ channel, the maximum energy is a little smaller by the same criterion, $T_\pi^\text{max}=32.7\MeV$, corresponding to 10 experiments and 180 data points. 
Finally in the charge-exchange $\pi^-p\to\pi^0 n$ channel, $T_\pi^\text{max}=45.6\MeV$, which includes 24 different experiments and 151 data points. 
We checked that the results do not depend significantly on the maximum energy considered, so that they are actually stable in a larger energy interval and the exact cutoff criterion becomes irrelevant. 
Note that we consider an independent normalization constant $\zeta$ for each experiment, 
in~\eqref{fit_strategy} this corresponds to $N=32$, $10$, and $24$ normalizations for $\pi^+p\to\pi^+p$, $\pi^-p\to\pi^-p$, and $\pi^-p\to\pi^0n$, respectively.

\begin{table*}[t!]
  \centering
  \renewcommand{\arraystretch}{1.3}
  \begin{tabular}{crrrrrr}
  \toprule
  & $\pi^+p\to\pi^+p$ &$\pi^-p\to\pi^-p$ &$\pi^-p\to\pi^0n$
\\\midrule
$a \,[10^{-3}\mpi^{-1}]$ & $-84.4(1.5)$ & $82.5(1.5)$ & $-122.3(3.4)$\\
$\chi^2/\text{dof}$ & $0.88$ & $0.79$ & $0.62$\\\midrule
$a \,[10^{-3}\mpi^{-1}]$ & $-80.7(1.5)$ & $83.6(1.5)$ & $-120.0(3.3)$\\
$\chi^2/\text{dof}$ & $0.95$ & $0.82$ & $0.67$\\\midrule
$a \,[10^{-3}\mpi^{-1}]$ & $-81.8(1.2)$ & $80.4(1.3)$ & \\
$\chi^2/\text{dof}$ & $1.38$ & $1.61$ & \\
   \bottomrule
  \end{tabular}
  \caption{Scattering lengths and reduced $\chi^2$ from the fit. First panel: full electromagnetic corrections from~\cite{Tromborg:1976bi}. Second panel: Coulomb corrections only.
  Third panel: as first panel, excluding data from~\cite{Denz:2005jq}.}
\label{tab:scattering_lengths}
\end{table*}

The iterative procedure described above converges reasonably fast. Defining a tolerance $\tau=\vert a^{i}-a^{i-1}\vert$ at each iteration $i$, one obtains stable minima corresponding to $\tau < 10^{-6}$ for the $\pi^-p\to\pi^-p$ and charge-exchange channels in less than 10 iterations. Due to larger numbers of normalization constants and data points, the convergence for $\pi^+p\to\pi^+p$ is slower, we obtain a tolerance $\tau< 10^{-3}$ in 50 iterations. 
The initial guesses for $a$, $\zzeta$, and $\Delta\zzeta$ in~\eqref{fit_strategy} are taken from~\eqref{scatt_length_pionic} and the GW solution~\cite{Workman:2012hx,SAID}, although the final solutions are stable irrespective of the considered starting values. In particular, the results are identical starting from the KH80 scattering lengths and/or $\zeta_k=1$.
This is an important cross check given the large parameter space of the fits. 
We stress that the self-consistent, unbiased determination of the normalizations according to~\eqref{fit_strategy} is crucial, in particular in the charge-exchange reaction, where otherwise an implausibly large departure from the isospin limit could be observed. 

The fit results are shown in Table~\ref{tab:scattering_lengths}: for all channels we obtain a good $\chi^2$. Switching off the non-Coulomb electromagnetic corrections, the description becomes slightly worse, but the fit values for the scattering lengths are rather stable. The biggest shift occurs in the $\pi^+p\to\pi^+p$ channel, whose scattering length is reduced by $2.5$ times its statistical uncertainty. 

\begin{table*}[p!]
{\small
  \centering
  \renewcommand{\arraystretch}{1.3}
  \begin{tabular}{rrrrrrrrr}
  \toprule
  \multicolumn{3}{c}{$\pi^+p\to\pi^+p$} &\multicolumn{3}{c}{$\pi^-p\to\pi^-p$} &\multicolumn{3}{c}{$\pi^-p\to\pi^0n$}\\
  Ref. &$\zeta_k$&$\zeta_k^\text{GW}$&
  Ref. &$\zeta_k$&$\zeta_k^\text{GW}$&
  Ref. &$\zeta_k$&$\zeta_k^\text{GW}$
\\\midrule
\cite{Denz:2005jq} & $0.93(5)$ & $0.94(4)$ & \cite{Denz:2005jq} & $1.00(6)$ & $1.01(5)$ & \cite{Isenhower:1999aj} & $0.81(14)$ & $0.92(3)$ \\
\cite{Denz:2005jq} & $1.07(7)$ & $1.04(5)$ & \cite{Denz:2005jq} & $1.13(10)$ & $1.19(7)$ & \cite{Isenhower:1999aj} & $0.79(18)$ & $0.96(4)$ \\
\cite{Bertin:1976uh} & $0.78(24)$ & $0.82(67)$ & \cite{Denz:2005jq} & $0.99(5)$ & $1.00(5)$ & \cite{Isenhower:1999aj} & $0.70(22)$ & $0.95(5)$ \\
\cite{Denz:2005jq} & $0.96(5)$ & $0.94(4)$ & \cite{Denz:2005jq} & $1.02(9)$ & $1.09(6)$ & \cite{Isenhower:1999aj} & $0.88(13)$ & $0.93(3)$ \\
\cite{Denz:2005jq} & $0.90(5)$ & $0.94(4)$ & \cite{Frank:1983ic} & $1.01(22)$ & $1.11(1.24)$ & \cite{Isenhower:1999aj} & $0.91(19)$ & $0.98(4)$ \\
\cite{Denz:2005jq} & $1.28(9)$ & $1.32(9)$ & \cite{Brack:1989sj} & $0.99(41)$ & $1.03(2)$ & \cite{Duclos:1973nb} & $1.10(26)$ & $1.00(0)$ \\
\cite{Frank:1983ic} & $1.10(17)$ & $1.16(1.35)$ & \cite{Denz:2005jq} & $1.05(24)$ & $1.06(6)$ & \cite{Salomon:1983xn} & $1.02(16)$ & $1.01(3)$ \\
\cite{Brack:1989sj} & $1.00(55)$ & $1.02(4)$ & \cite{Denz:2005jq} & $1.01(8)$ & $1.05(6)$ & \cite{Frlez:1997qu} & $0.77(32)$ & $0.96(8)$ \\
\cite{Bertin:1976uh} & $0.86(25)$ & $0.90(81)$ & \cite{Joram:1995gr} & $0.98(5)$ & $1.00(3)$ & \cite{Fitzgerald:1986fg} & $0.74(56)$ & $1.04(9)$ \\
\cite{Denz:2005jq} & $0.89(47)$ & $0.92(4)$ & \cite{Joram:1995gr} & $1.05(51)$ & $1.08(4)$ & \cite{Duclos:1973nb} &  $1.00(20)$ & $1.00(0)$ \\
\cite{Denz:2005jq} & $0.89(5)$ & $0.94(4)$ &     & & & \cite{Mekterovic:2009kw} & $0.93(12)$ & $0.95(3)$ \\
\cite{Denz:2005jq} & $1.01(6)$ & $1.06(6)$ &     & & & \cite{Jia:2008rt} & $1.12(80)$ & $1.39(19)$ \\
\cite{Joram:1995gr} & $1.05(6)$ & $1.06(4)$ &     & & & \cite{Fitzgerald:1986fg} & $0.70(57)$ & $1.04(9)$ \\
\cite{Joram:1995gr} & $1.15(11)$ & $1.03(4)$ &    & & & \cite{Salomon:1983xn} & $0.98(14)$ & $0.99(3)$ \\
\cite{Denz:2005jq} & $0.95(16)$ & $1.01(8)$ &    & & & \cite{Isenhower:1999aj} & $1.00(12)$ & $0.99(4)$ \\
\cite{Denz:2005jq} & $1.05(9)$ & $1.10(10)$ &    & & & \cite{Isenhower:1999aj} & $0.88(16)$ & $0.93(3)$ \\
\cite{Denz:2005jq} & $1.12(10)$ & $1.18(11)$ &   & & & \cite{Isenhower:1999aj} & $0.81(56)$ & $1.06(7)$ \\
\cite{Bertin:1976uh} & $0.92(38)$ & $0.92(84)$ &   & & & \cite{Mekterovic:2009kw} & $0.95(11)$ & $0.96(2)$ \\
\cite{Blecher:1979zz} & $0.81(4)$ & $0.96(4)$ &     & & & \cite{Jia:2008rt} & $1.23(78)$ & $1.37(19)$ \\
\cite{Denz:2005jq} & $0.66(10)$ & $0.84(5)$ &    & & & \cite{Fitzgerald:1986fg} & $0.71(78)$ & $1.04(9)$ \\
\cite{Denz:2005jq} & $0.99(7)$ & $1.01(7)$ &     & & & \cite{Duclos:1973nb} & $1.14(20)$ & $1.00(0)$ \\
\cite{Denz:2005jq} & $1.02(7)$ & $1.09(8)$ &     & & & \cite{Jia:2008rt} & $1.29(72)$ & $1.30(17)$ \\
\cite{Denz:2005jq} & $0.82(10)$ & $0.87(5)$ &    & & & \cite{Mekterovic:2009kw} & $0.95(11)$ & $0.97(3)$ \\
\cite{Denz:2005jq} & $0.92(6)$ & $0.95(6)$ &     & & & \cite{Bagheri:1987kc} & $0.95(11)$ & $0.96(3)$ \\
\cite{Denz:2005jq} & $1.03(8)$ & $1.09(8)$ &     & & & & & \\								
\cite{Joram:1995gr} & $1.08(8)$ & $1.12(4)$ &     & & & & & \\								
\cite{Brack:1989sj} & $1.02(4)$ & $1.06(3)$ &     & & & & & \\								
\cite{Joram:1995gr} & $1.18(6)$ & $1.12(4)$ &     & & & & & \\								
\cite{Auld:1979yd}  & $0.99(20)$ & $1.05(1.11)$ & & & & & & \\								
\cite{Frank:1983ic} & $0.99(13)$ & $1.01(1.02)$ & & & & & & \\								
\cite{Moinester:1978zu} & $0.90(22)$ & $0.93(6)$ &    & & & & & \\								
\cite{Bertin:1976uh} & $0.90(25)$ & $0.94(87)$ &   & & & & & \\
   \bottomrule
  \end{tabular}
}
  \caption{Fit normalization constants $\zeta_k$ for each experiment included in the analysis, compared to the GW normalizations $\zeta_k^\text{GW}$. 
    Note that the $\zeta_k^\text{GW}$ correspond to the inverse of the normalizations provided in the GW data base.}
\label{tab:normalizations}
\end{table*}

The results for the normalizations are given in Table~\ref{tab:normalizations}, where we also provide the GW values~\cite{Workman:2012hx,SAID} for comparison. The central values agree within uncertainties in all three channels, but the uncertainties differ substantially in some cases. Our uncertainty estimates, derived from the Hessian at the fit minimum, do not produce the very large $\Order(1)$ uncertainties as observed in the GW fit. However, especially for the charge-exchange reaction, some normalizations still carry sizable uncertainties, which may be related to inconsistencies in the data base or underestimated experimental errors already noticed in~\cite{Mekterovic:2009kw,Jia:2008rt}.

\begin{figure}
\centering
\includegraphics[width=0.94\linewidth,clip]{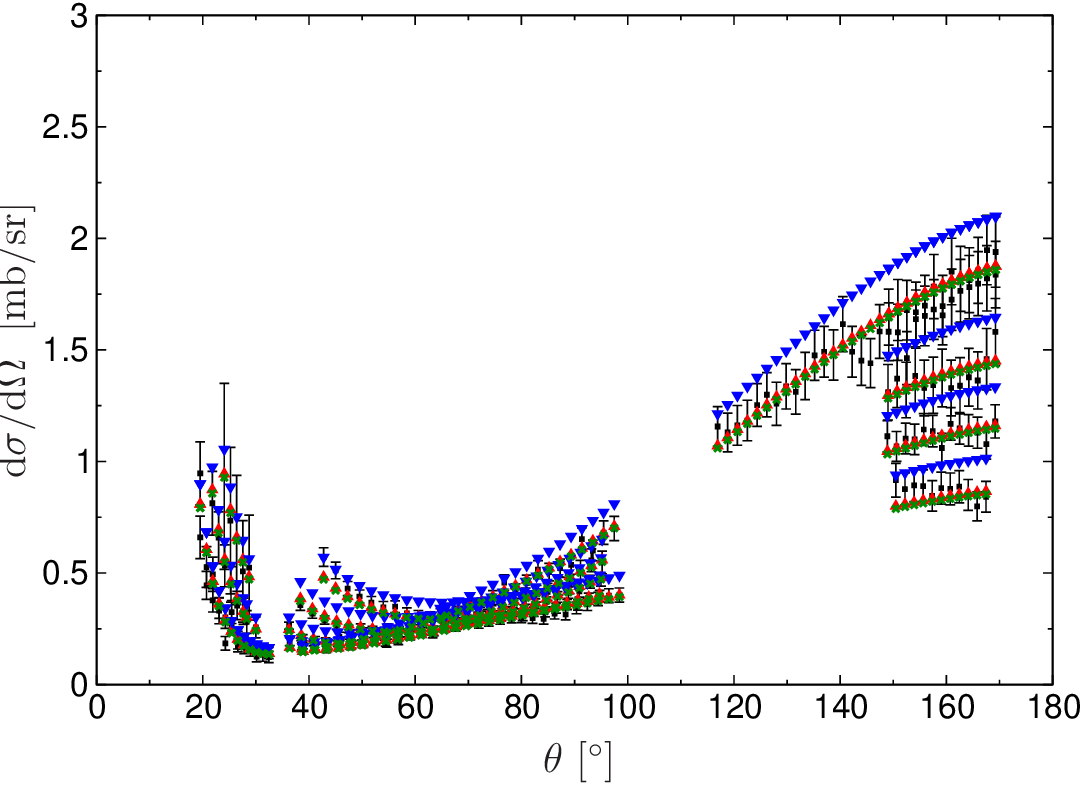}\\
\includegraphics[width=0.94\linewidth,clip]{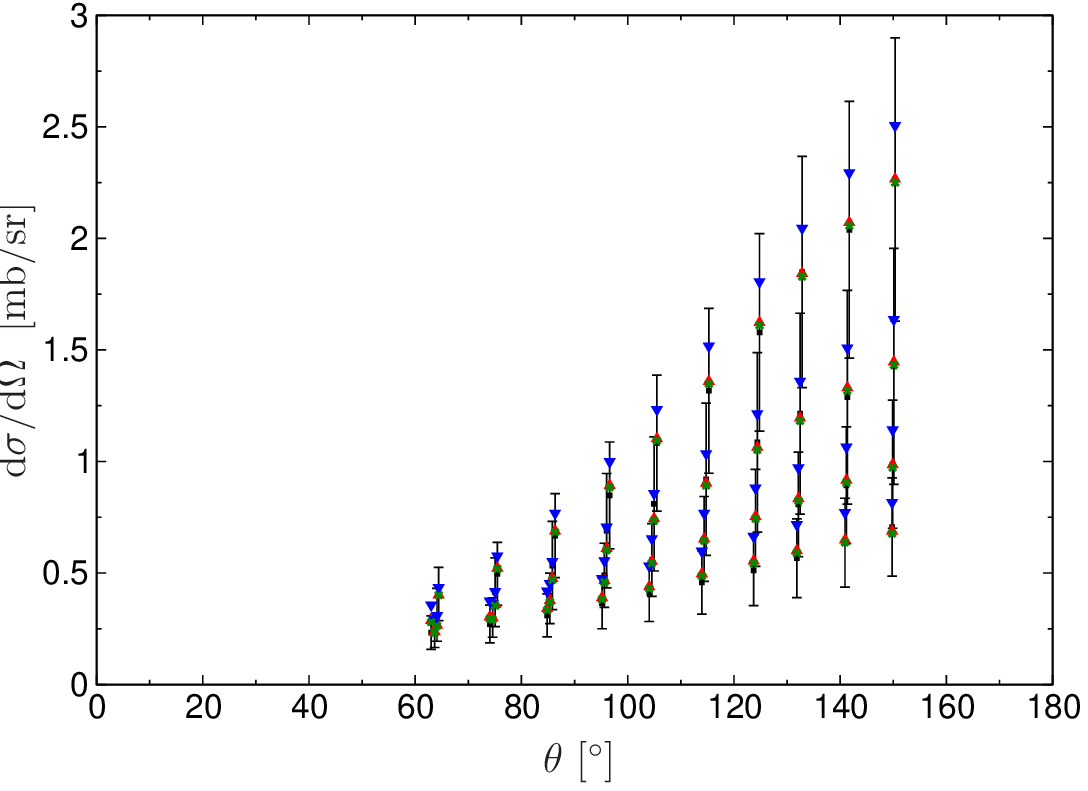}
\caption{$\pi^+ p\to \pi^+ p$ differential cross section as a function of the scattering angle $\theta$ for $T_\pi \leq T_\pi^{\text{max}}$. 
The experimental data (black error bars) are taken from the SAID/GWU data base~\cite{Workman:2012hx,SAID} with normalizations $\zeta_k$ from Table~\ref{tab:normalizations} extracted from the data fit; the two panels include the data from \cite{Denz:2005jq} (top) and \cite{Bertin:1976uh} (bottom).
Green crosses refer to the RS representation in~\eqref{eq:dcross} with the scattering lengths from the fit solution in Table~\ref{tab:scattering_lengths}.  
For comparison, we also include the RS solutions generated with scattering lengths from pionic atoms~\eqref{scatt_length_pionic} (red up triangles) and KH80~\eqref{KH80} (down blue triangles).}
\label{fig:pipDSC_1}
\end{figure}

\begin{figure}
\centering
\includegraphics[width=0.94\linewidth,clip]{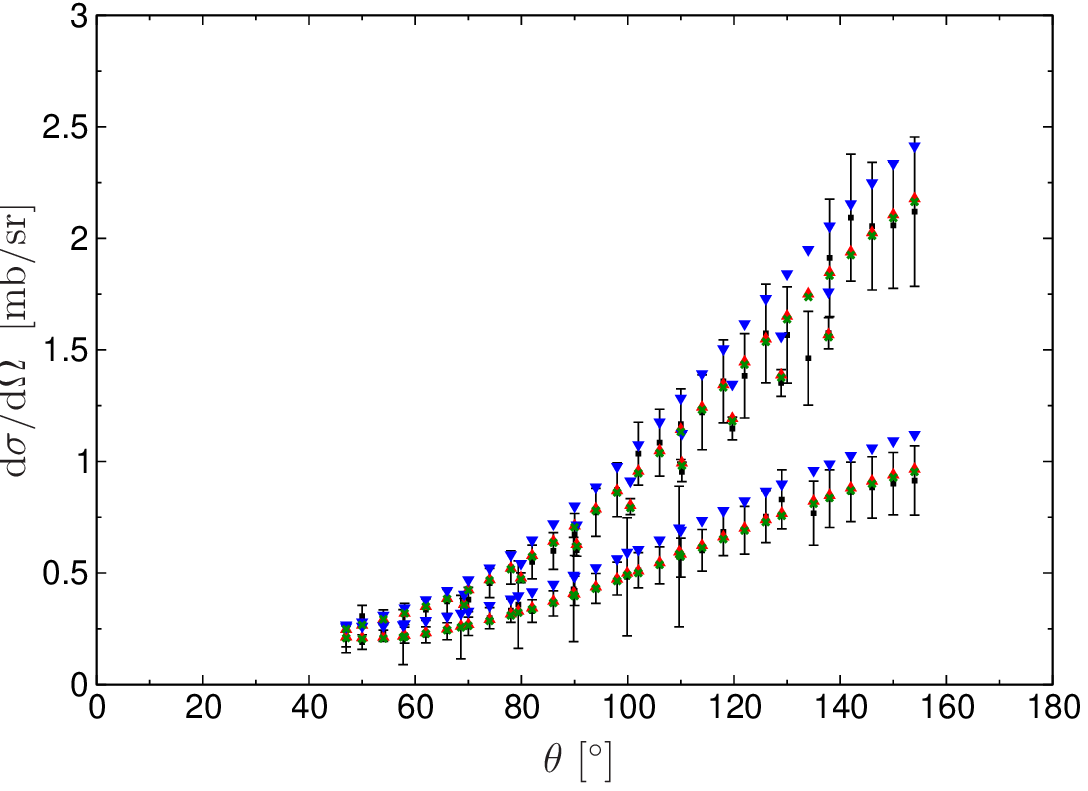}\\
\includegraphics[width=0.94\linewidth,clip]{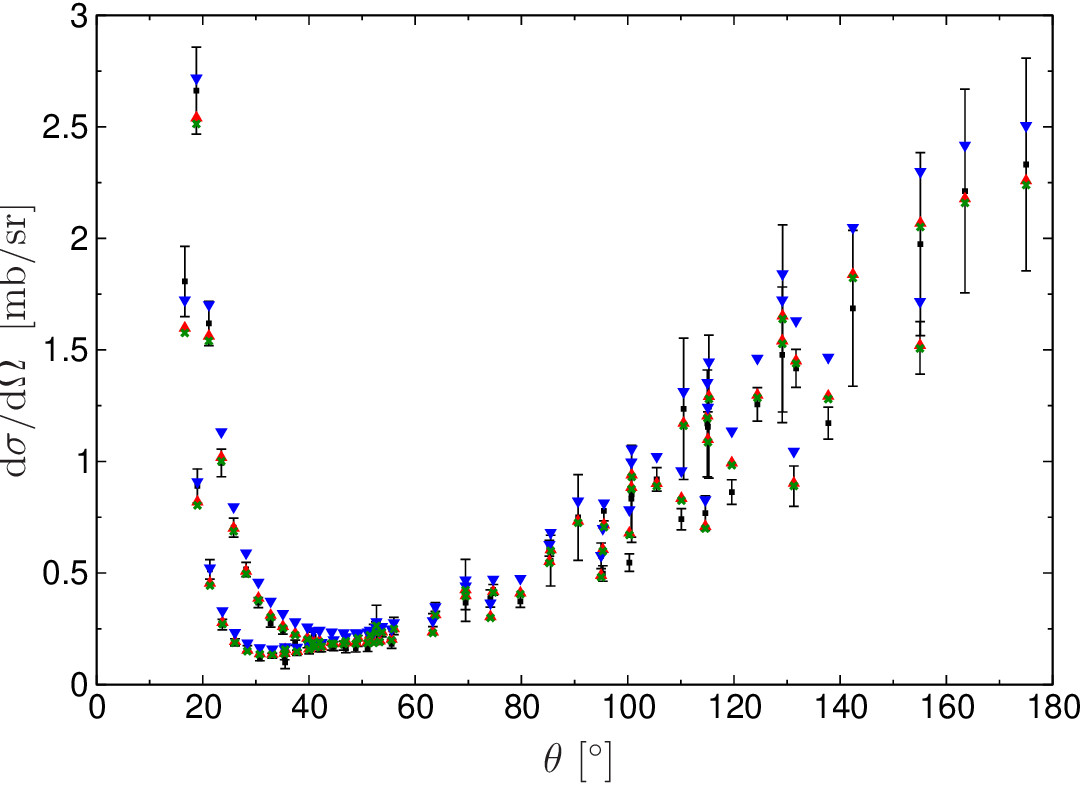}
\caption{$\pi^+ p\to \pi^+ p$ differential cross section as a function of the scattering angle $\theta$ for $T_\pi \leq T_\pi^{\text{max}}$, as in Fig.~\ref{fig:pipDSC_1}.
The experimental data (black error bars) are taken from the SAID/GWU data base~\cite{Workman:2012hx,SAID} with normalizations $\zeta_k$ from Table~\ref{tab:normalizations} extracted from the data fit; the two panels include the data from~\cite{Frank:1983ic,Brack:1989sj} (top) and \cite{Joram:1995gr,Blecher:1979zz,Auld:1979yd,Moinester:1978zu} (bottom).  
Green crosses refer to the RS representation in~\eqref{eq:dcross} with the scattering lengths from the fit solution in Table~\ref{tab:scattering_lengths}.  
For comparison, we also include the RS solutions generated with scattering lengths from pionic atoms~\eqref{scatt_length_pionic} (red up triangles) and KH80~\eqref{KH80} (down blue triangles).}
\label{fig:pipDSC_2}
\end{figure}

For illustration, we plot the $\pi^+ p\to \pi^+p$ differential cross section results in Figs.~\ref{fig:pipDSC_1} and \ref{fig:pipDSC_2}, as a function of the scattering angle. The black error bars correspond to the $N=32$ experiments  below $T_\pi^{\text{max}}$, where we include measurements for different energies within the same plot.
Green crosses denote the RS fit results, corresponding to the scattering lengths in Table~\ref{tab:scattering_lengths}. In addition, we also plot the RS results generated with pionic-atom and KH80 scattering lengths for comparison, denoted by red up triangles and blue down triangles, respectively.  
The figures reveal that already by eye the data are incompatible with the KH80 solution, while the pionic-atom results are in perfect agreement.  

Since the data base for the elastic channels is dominated by~\cite{Denz:2005jq}, we also performed fits excluding these data sets, with results shown in the third panel of Table~\ref{tab:scattering_lengths} (for both channels~\cite{Denz:2005jq} accounts for more than half the data points, $274/464$ and $121/180$ for $\pi^\pm p$). The pull from the full data base amounts to $1.4\sigma$ and $1.1\sigma$ for the $\pi^+p$ and $\pi^-p$ channel, respectively, and especially given that the statistical errors for the reduced fit set should be inflated due to the increased $\chi^2$, we conclude that the results for the full and reduced data base are compatible even within statistical errors only, both for the scattering lengths and the normalizations, while possible inconsistencies between the data sets do not become relevant at the level of accuracy for the scattering lengths that we are claiming here. For this reason, we continue to use the full data base in the following. 

Adding the statistical uncertainties, the uncertainties related to the non-Coulomb electromagnetic corrections, and the potential shifts from virtual photons in ChPT~\eqref{ChPT_virtual} in quadrature, we obtain
\begin{align}
\label{scatt_length_fit}
 a_{\pi^+p\to\pi^+p}&=-84.4(4.2)\times 10^{-3}\mpi^{-1},\notag\\
 a_{\pi^-p\to\pi^-p}&=82.5(2.1)\times 10^{-3}\mpi^{-1},\notag\\
 a_{\pi^-p\to\pi^0n}&=-122.3(4.1)\times 10^{-3}\mpi^{-1}.
\end{align}
With three channels but only two independent amplitudes in the isospin limit, we can quantify the amount of isospin breaking in terms of the triangle relation
\beq
\label{R}
R=2\frac{a_{\pi^+p\to\pi^+p}-a_{\pi^-p\to\pi^-p}-\sqrt{2}a_{\pi^-p\to\pi^0n}}{a_{\pi^+p\to\pi^+p}-a_{\pi^-p\to\pi^-p}+\sqrt{2}a_{\pi^-p\to\pi^0n}}
=-3.6(4.4)\%
\eeq
directly from experiment, to be compared with the chiral prediction $R=0.6(4)\%$ (updating~\cite{Hoferichter:2009ez} to remove the effects from virtual photons).
The experimental result is therefore consistent with the small amount of isospin violation as predicted by ChPT (see~\cite{Meissner:1997ii,Fettes:1998wf,Fettes:2000vm,Fettes:2001cr} for earlier analyses) and found within the $K$-matrix model of~\cite{Gridnev:2004mk}.
However, a conclusive test of the large amount of isospin violation $R\sim-7\%$ found in earlier phenomenological models~\cite{Gibbs:1995dm,Matsinos:1997pb}
would require better control over radiative corrections.

\section{Consequences for the pion--nucleon $\boldsymbol{\sigma}$-term}
\label{sec:sigma_term}

To derive the $\sigma$-term corresponding to the scattering lengths~\eqref{scatt_length_fit} we need to convert the results to the isospin basis (defined by the elastic reactions). For this reason, there is another isospin-breaking correction $\Delta a_{\pi^-p\to\pi^0n}=0.8(0.5)\times 10^{-3}\mpi^{-1}$ that needs to be subtracted from $a_{\pi^-p\to\pi^0n}$, essentially corresponding to the ChPT prediction for the triangle relation $R$, but in view of the uncertainties this is a very minor effect. 

\begin{figure}
\centering
\includegraphics[width=\linewidth]{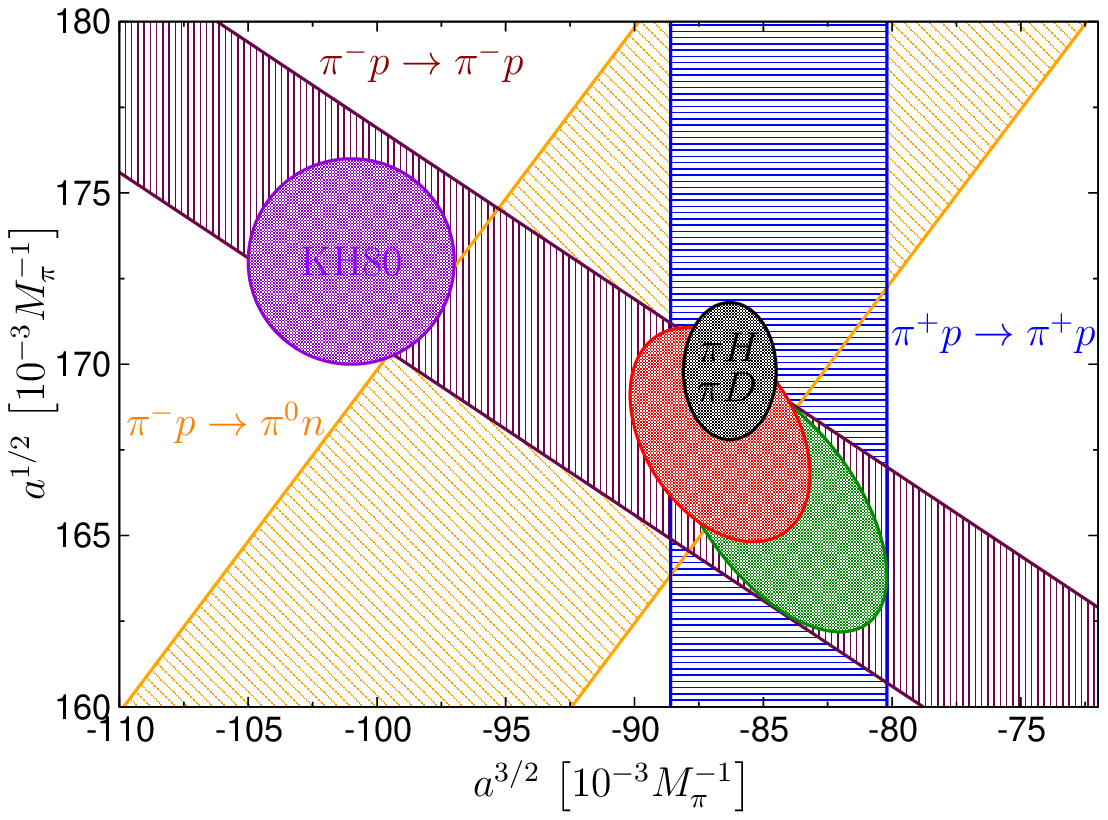}
\caption{Constraints on $a^{1/2}$ and $a^{3/2}$ from $\pi^+p\to\pi^+p$ (blue), $\pi^-p\to\pi^-p$ (maroon), and $\pi^-p\to\pi^0 n$ (orange). The combination of the elastic (all) channels leads to the green (red) ellipse. The KH80 and pionic-atom scattering lengths are marked in violet and black, respectively.}
\label{fig:bands}
\end{figure}

First, it is instructive to compare the fit results~\eqref{scatt_length_fit} with the expectation from the pionic-atom or KH80 scattering lengths
\begin{align}
 \bar a_{\pi^+p\to\pi^+p}&=-86.3(1.8)\times 10^{-3}\mpi^{-1},& a_{\pi^+p\to\pi^+p}^\text{KH80}&=-101(4)\times 10^{-3}\mpi^{-1},\notag\\
 \bar a_{\pi^-p\to\pi^-p}&=84.4(1.5)\times 10^{-3}\mpi^{-1}, & a_{\pi^-p\to\pi^-p}^\text{KH80}&=81.7(2.4)\times 10^{-3}\mpi^{-1},\notag\\
 \bar a_{\pi^-p\to\pi^0n}&=-120.7(1.3)\times 10^{-3}\mpi^{-1},& a_{\pi^-p\to\pi^0n}^\text{KH80}&=-129.2(2.4)\times 10^{-3}\mpi^{-1}.
\end{align}
While the $\pi^-p\to\pi^-p$ data are thus compatible with both, the charge-exchange reaction displays a mild preference for the pionic-atom values, and the $\pi^+p\to\pi^+p$ data strongly disfavor the KH80 scattering length. As expected, it is the latter reaction that allows us to discriminate between the two sets. 
This observation is illustrated in Fig.~\ref{fig:bands}: while the bands in the $a^{1/2}$--$a^{3/2}$ plane from $\pi^-p\to\pi^-p$ and $\pi^-p\to\pi^0n$ are not conclusive, the constraint from $\pi^+p\to\pi^+p$ decides in favor of pionic atoms.   

The resulting scattering lengths from the combination of the three reactions, either from the elastic reactions only (green ellipse) or all channels (red ellipse), are quoted in Table~\ref{tab:sigma_term}, together with the corresponding value of $\sigma_{\pi N}$ derived from~\eqref{sigma_piN_lin}. Given that the charge-exchange data are fully consistent with the elastic reactions, with isospin-breaking effects compatible with zero at the level of~\eqref{R}, we adopt the combination of all channels as our final result and quote
\beq
\sigma_{\pi N}=58(5)\MeV
\eeq
for the $\pi N$ $\sigma$-term, directly derived from $\pi N$ scattering data. This value is fully consistent with the result from pionic atoms~\cite{Hoferichter:2015dsa}, and provides further evidence
that the discrepancy with lattice calculations cannot be blamed on deficient $\pi N$ data input. 

\begin{table*}[t!]
  \centering
  \renewcommand{\arraystretch}{1.3}
  \begin{tabular}{crrrr}
  \toprule
  & $a^{1/2}\, [10^{-3}\mpi^{-1}]$ & $a^{3/2}\, [10^{-3}\mpi^{-1}]$ & $\rho$ & $\sigma_{\pi N}\, [\text{MeV}]$
\\\midrule
all channels & $-86.7(3.5)$ & $167.9(3.2)$ & $-0.36$ & $58.3(4.2)$\\
only $\pi^\pm p\to\pi^\pm p$ & $-84.4(4.2)$ & $166.0(3.8)$ & $-0.55$ & $59.8(4.5)$\\
   \bottomrule
  \end{tabular}
  \caption{Scattering lengths in the isospin basis including the correlation coefficients $\rho$ and the corresponding values for $\sigma_{\pi N}$.}
\label{tab:sigma_term}
\end{table*}

\section{Conclusions}
\label{sec:con}

The direct extraction of the $\pi N$ scattering lengths from the low-energy data base, given in~\eqref{scatt_length_fit}, constitutes the main result of the present paper.
Even after accounting for the uncertainties introduced by electromagnetic corrections as well as the normalizations in each experiment, the resulting constraint is sufficiently
precise to conclusively test the scattering lengths from pionic-atom data against earlier extractions from $\pi N$ scattering~\cite{Koch:1980ay,Hoehler}. Remarkably, the three measured scattering channels
lead to a consistent picture that fully confirms the pionic-atom results, and, at the present level of accuracy, does not exhibit evidence for isospin violation, 
in turn confirming expectations from chiral perturbation theory.
Arguably, these results raise the status of the $\pi N$ scattering lengths to a similar level as for $\pi\pi$ scattering~\cite{Leutwyler:2015jga}, with Roy-equation-based extractions from different experiments 
in perfect agreement.

A direct consequence concerns the phenomenological value of the $\pi N$ $\sigma$-term, whose determination crucially depends on the scattering length input. Confirming the pionic-atom extraction from another experimental source, albeit with larger uncertainties, thereby provides an independent verification of~\cite{Hoferichter:2015dsa} and makes it appear unlikely that the $\pi N$ scattering lengths are the
origin of the current tension with lattice QCD.    

\section*{Acknowledgments}

Financial support by
the DFG (SFB/TR 16, ``Subnuclear Structure of Matter,'' CRC 110, ``Symmetries and the Emergence of Structure
in QCD''),  
the DOE (Grant No.\ DE-FG02-00ER41132), 
and the Swiss National Science Foundation
is gratefully acknowledged.
The work of UGM was supported in part by The Chinese Academy of Sciences 
(CAS) President's International Fellowship Initiative (PIFI) Grant No.\ 2017VMA0025.


\section*{References}

\begin{thebibliography}{99}

\bibitem{Crivellin:2013ipa} 
  A.~Crivellin, M.~Hoferichter and M.~Procura,
  Phys.\ Rev.\ D {\bf 89} (2014) 054021
  [arXiv:1312.4951 [hep-ph]].
  
\bibitem{Bottino:1999ei} 
  A.~Bottino, F.~Donato, N.~Fornengo and S.~Scopel,
  Astropart.\ Phys.\  {\bf 13} (2000) 215
  [hep-ph/9909228].
  
\bibitem{Bottino:2001dj}
  A.~Bottino, F.~Donato, N.~Fornengo and S.~Scopel,
  Astropart.\ Phys.\  {\bf 18} (2002) 205
  [hep-ph/0111229].
  
\bibitem{Ellis:2008hf} 
  J.~R.~Ellis, K.~A.~Olive and C.~Savage,
  Phys.\ Rev.\ D {\bf 77} (2008) 065026
  [arXiv:0801.3656 [hep-ph]].
  
\bibitem{Cirigliano:2009bz}
  V.~Cirigliano, R.~Kitano, Y.~Okada and P.~Tuzon,
  Phys.\ Rev.\ D {\bf 80} (2009) 013002
  [arXiv:0904.0957 [hep-ph]].
  
\bibitem{Crivellin:2014cta} 
  A.~Crivellin, M.~Hoferichter and M.~Procura,
  Phys.\ Rev.\ D {\bf 89} (2014) 093024 
  [arXiv:1404.7134 [hep-ph]].
  
\bibitem{Engel:2013lsa}
  J.~Engel, M.~J.~Ramsey-Musolf and U.~van Kolck,
  Prog.\ Part.\ Nucl.\ Phys.\  {\bf 71} (2013) 21
  [arXiv:1303.2371 [nucl-th]].
  
\bibitem{deVries:2015gea}
  J.~de Vries and U.-G.~Mei{\ss}ner,
  Int.\ J.\ Mod.\ Phys.\ E {\bf 25} (2016)  1641008
  [arXiv:1509.07331 [hep-ph]].
  
\bibitem{deVries:2016jox}
  J.~de Vries, E.~Mereghetti, C.~Y.~Seng and A.~Walker-Loud,
  Phys.\ Lett.\ B {\bf 766} (2017) 254
  [arXiv:1612.01567 [hep-lat]].
  
\bibitem{Yamanaka:2017mef}
  N.~Yamanaka, B.~K.~Sahoo, N.~Yoshinaga, T.~Sato, K.~Asahi and B.~P.~Das,
  Eur.\ Phys.\ J.\ A {\bf 53} (2017) 54
  [arXiv:1703.01570 [hep-ph]].
  
\bibitem{Cheng:1970mx} 
  T.~P.~Cheng and R.~F.~Dashen,
  Phys.\ Rev.\ Lett.\  {\bf 26} (1971) 594.
  
\bibitem{Brown:1971pn} 
  L.~S.~Brown, W.~J.~Pardee and R.~D.~Peccei,
  Phys.\ Rev.\ D {\bf 4} (1971) 2801.
  
\bibitem{Gasser:1988jt} 
  J.~Gasser, H.~Leutwyler, M.~P.~Locher and M.~E.~Sainio,
  Phys.\ Lett.\ B {\bf 213} (1988) 85.
  
\bibitem{Gasser:1990ce} 
  J.~Gasser, H.~Leutwyler and M.~E.~Sainio,
  Phys.\ Lett.\ B {\bf 253} (1991) 252.
  
\bibitem{Gasser:1990ap} 
  J.~Gasser, H.~Leutwyler and M.~E.~Sainio,
  Phys.\ Lett.\ B {\bf 253} (1991) 260.
  
\bibitem{Koch:1980ay} 
  R.~Koch and E.~Pietarinen,
  Nucl.\ Phys.\ A {\bf 336} (1980) 331.
 
\bibitem{Hoehler}
  G.~H\"{o}hler,
  {\it Pion--Nukleon-Streuung: Methoden und Ergebnisse},
  in Landolt-B\"ornstein, {\bf 9b2}, ed.\ H.~Schopper,
  Springer Verlag, Berlin, 1983.  
  
\bibitem{Pavan:2001wz} 
  M.~M.~Pavan, I.~I.~Strakovsky, R.~L.~Workman and R.~A.~Arndt,
  $\pi N$ Newslett.\  {\bf 16} (2002) 110
  [hep-ph/0111066].

\bibitem{Ditsche:2012fv}
  C.~Ditsche, M.~Hoferichter, B.~Kubis and U.-G.~Mei{\ss}ner,
  JHEP {\bf 1206} (2012) 043
  [arXiv:1203.4758 [hep-ph]].

\bibitem{Hoferichter:2012wf}
  M.~Hoferichter, C.~Ditsche, B.~Kubis and U.-G.~Mei{\ss}ner,
  JHEP {\bf 1206} (2012) 063
  [arXiv:1204.6251 [hep-ph]].

\bibitem{Hoferichter:2015dsa}
  M.~Hoferichter, J.~Ruiz de Elvira, B.~Kubis and U.-G.~Mei{\ss}ner,
  Phys.\ Rev.\ Lett.\  {\bf 115} (2015) 092301
  [arXiv:1506.04142 [hep-ph]].

\bibitem{Hoferichter:2015tha}
  M.~Hoferichter, J.~Ruiz de Elvira, B.~Kubis and U.-G.~Mei{\ss}ner,
  Phys.\ Rev.\ Lett.\  {\bf 115} (2015)  192301
  [arXiv:1507.07552 [nucl-th]].

\bibitem{Hoferichter:2015hva}
  M.~Hoferichter, J.~Ruiz de Elvira, B.~Kubis and U.-G.~Mei{\ss}ner,
  Phys.\ Rept.\  {\bf 625} (2016) 1
  [arXiv:1510.06039 [hep-ph]].

\bibitem{Hoferichter:2016ocj}
  M.~Hoferichter, J.~Ruiz de Elvira, B.~Kubis and U.-G.~Mei{\ss}ner,
  Phys.\ Lett.\ B {\bf 760} (2016) 74
  [arXiv:1602.07688 [hep-lat]].

\bibitem{Hoferichter:2016duk}
  M.~Hoferichter, B.~Kubis, J.~Ruiz de Elvira, H.-W.~Hammer and U.-G.~Mei{\ss}ner,
  Eur.\ Phys.\ J.\ A {\bf 52} (2016)  331
  [arXiv:1609.06722 [hep-ph]].

\bibitem{Siemens:2016jwj}
  D.~Siemens, J.~Ruiz de Elvira, E.~Epelbaum, M.~Hoferichter, H.~Krebs, B.~Kubis and U.-G.~Mei{\ss}ner,
  Phys.\ Lett.\ B {\bf 770} (2017) 27
  [arXiv:1610.08978 [nucl-th]].
  
\bibitem{Strauch:2010vu} 
  T.~Strauch {\it et al.},
  Eur.\ Phys.\ J.\ A {\bf 47} (2011) 88
  [arXiv:1011.2415 [nucl-ex]].
  
\bibitem{Hennebach:2014lsa} 
  M.~Hennebach {\it et al.},
  Eur.\ Phys.\ J.\ A {\bf 50} (2014) 190
  [arXiv:1406.6525 [nucl-ex]].
  
\bibitem{Gotta:2008zza} 
  D.~Gotta {\it et al.},
  Lect.\ Notes Phys.\  {\bf 745} (2008) 165.
  
\bibitem{Baru:2010xn}
  V.~Baru, C.~Hanhart, M.~Hoferichter, B.~Kubis, A.~Nogga and D.~R.~Phillips,
  Phys.\ Lett.\ B {\bf 694} (2011) 473
  [arXiv:1003.4444 [nucl-th]].
  
\bibitem{Baru:2011bw}
  V.~Baru, C.~Hanhart, M.~Hoferichter, B.~Kubis, A.~Nogga and D.~R.~Phillips,
  Nucl.\ Phys.\ A {\bf 872} (2011) 69
  [arXiv:1107.5509 [nucl-th]].

\bibitem{Gasser:2002am}
  J.~Gasser, M.~A.~Ivanov, E.~Lipartia, M.~Moj\v zi\v s and A.~Rusetsky,
  Eur.\ Phys.\ J.\ C {\bf 26} (2002) 13
  [hep-ph/0206068].
  
\bibitem{Hoferichter:2009ez}
  M.~Hoferichter, B.~Kubis and U.-G.~Mei{\ss}ner,
  Phys.\ Lett.\ B {\bf 678} (2009) 65
  [arXiv:0903.3890 [hep-ph]].
  
\bibitem{Hoferichter:2009gn}
  M.~Hoferichter, B.~Kubis and U.-G.~Mei{\ss}ner,
  Nucl.\ Phys.\ A {\bf 833} (2010) 18
  [arXiv:0909.4390 [hep-ph]].
  
\bibitem{Hoferichter:2012bz}
  M.~Hoferichter, V.~Baru, C.~Hanhart, B.~Kubis, A.~Nogga and D.~R.~Phillips,
  PoS CD {\bf 12} (2013) 093
  [arXiv:1211.1145 [nucl-th]].
  
\bibitem{Fettes:2000xg}
  N.~Fettes and U.-G.~Mei{\ss}ner,
  Nucl.\ Phys.\ A {\bf 676} (2000) 311
  [hep-ph/0002162].
  
\bibitem{Alarcon:2011zs}
  J.~M.~Alarc\'on, J.~Martin Camalich and J.~A.~Oller,
  Phys.\ Rev.\ D {\bf 85} (2012) 051503
  [arXiv:1110.3797 [hep-ph]].
  
\bibitem{Durr:2015dna}
  S.~D\"urr {\it et al.} [BMW Collaboration],
  Phys.\ Rev.\ Lett.\  {\bf 116} (2016) 172001
  [arXiv:1510.08013 [hep-lat]].
  
\bibitem{Yang:2015uis}
  Y.~B.~Yang {\it et al.} [$\chi$QCD Collaboration],
  Phys.\ Rev.\ D {\bf 94} (2016) 054503
  [arXiv:1511.09089 [hep-lat]].
  
\bibitem{Abdel-Rehim:2016won}
  A.~Abdel-Rehim {\it et al.} [ETM Collaboration],
  Phys.\ Rev.\ Lett.\  {\bf 116} (2016)  252001
  [arXiv:1601.01624 [hep-lat]].
  
\bibitem{Bali:2016lvx}
  G.~S.~Bali {\it et al.} [RQCD Collaboration],
  Phys.\ Rev.\ D {\bf 93} (2016) 094504
  [arXiv:1603.00827 [hep-lat]].
 
\bibitem{Wendt:2014lja}
  K.~A.~Wendt, B.~D.~Carlsson and A.~Ekstr\"om,
  arXiv:1410.0646 [nucl-th].
  
\bibitem{Siemens:2016hdi}
  D.~Siemens, V.~Bernard, E.~Epelbaum, A.~Gasparyan, H.~Krebs and U.-G.~Mei{\ss}ner,
  Phys.\ Rev.\ C {\bf 94} (2016)  014620
  [arXiv:1602.02640 [nucl-th]].
  
\bibitem{Siemens:2017opr}
  D.~Siemens, V.~Bernard, E.~Epelbaum, A.~M.~Gasparyan, H.~Krebs and U.-G.~Mei{\ss}ner,
  arXiv:1704.08988 [nucl-th].
  
\bibitem{Workman:2012hx}
  R.~L.~Workman, R.~A.~Arndt, W.~J.~Briscoe, M.~W.~Paris and I.~I.~Strakovsky,
  Phys.\ Rev.\ C {\bf 86} (2012) 035202
  [arXiv:1204.2277 [hep-ph]].
  
\bibitem{SAID}
  SAID, \texttt{http://gwdac.phys.gwu.edu/analysis/pin\_analysis.html}.  

\bibitem{Matsinos:2016fcd}
  E.~Matsinos and G.~Rasche,
  Int.\ J.\ Mod.\ Phys.\ E {\bf 26} (2017)  1750002
  [arXiv:1610.02921 [nucl-th]].
 
\bibitem{Tromborg:1973zt}
  B.~Tromborg and J.~Hamilton,
  Nucl.\ Phys.\ B {\bf 76} (1974) 483.
  
\bibitem{Tromborg:1975mn}
  B.~Tromborg, S.~Waldenstr\o m and I.~\O verb\o,
  Annals Phys.\  {\bf 100} (1976) 1.
  
\bibitem{Tromborg:1976bi}
  B.~Tromborg, S.~Waldenstr\o m and I.~\O verb\o,
  Phys.\ Rev.\ D {\bf 15} (1977) 725.
  
\bibitem{Tromborg:1977da}
  B.~Tromborg, S.~Waldenstr\o m and I.~\O verb\o,
  Helv.\ Phys.\ Acta {\bf 51} (1978) 584.
 
\bibitem{Ananthanarayan:2000ht}
  B.~Ananthanarayan, G.~Colangelo, J.~Gasser and H.~Leutwyler,
  Phys.\ Rept.\  {\bf 353} (2001) 207
  [hep-ph/0005297].
  
\bibitem{Colangelo:2001df}
  G.~Colangelo, J.~Gasser and H.~Leutwyler,
  Nucl.\ Phys.\ B {\bf 603} (2001) 125
  [hep-ph/0103088].
  
\bibitem{Colangelo:2001sp}
  G.~Colangelo, J.~Gasser and H.~Leutwyler,
  Phys.\ Rev.\ Lett.\  {\bf 86} (2001) 5008
  [hep-ph/0103063].

\bibitem{GarciaMartin:2011cn}
  R.~Garc\'ia-Mart\'in, R.~Kami\'nski, J.~R.~Pel\'aez, J.~Ruiz de Elvira, and F.~J.~Yndur\'ain,
  Phys.\ Rev.\  D {\bf 83} (2011)  074004
  [arXiv:1102.2183 [hep-ph]].
 
\bibitem{DAgostini:1993arp}
  G.~D'Agostini,
  Nucl.\ Instrum.\ Meth.\ A {\bf 346} (1994) 306.
 
\bibitem{Ball:2009qv}
  R.~D.~Ball {\it et al.} [NNPDF Collaboration],
  JHEP {\bf 1005} (2010) 075
  [arXiv:0912.2276 [hep-ph]].

\bibitem{Denz:2005jq} 
  H.~Denz {\it et al.},
  Phys.\ Lett.\ B {\bf 633} (2006) 209
  [nucl-ex/0512006].

\bibitem{Isenhower:1999aj} 
  L.~D.~Isenhower {\it et al.},
  $\pi N$ Newslett.\  {\bf 15} (1999) 292.
   
\bibitem{Bertin:1976uh} 
  P.~Y.~Bertin {\it et al.},
  Nucl.\ Phys.\ B {\bf 106} (1976) 341.

\bibitem{Frank:1983ic} 
  J.~S.~Frank {\it et al.},
  Phys.\ Rev.\ D {\bf 28} (1983) 1569.

\bibitem{Brack:1989sj} 
  J.~T.~Brack {\it et al.},
  Phys.\ Rev.\ C {\bf 41} (1990) 2202.

\bibitem{Duclos:1973nb} 
  J.~Duclos {\it et al.},
  Phys.\ Lett.\  {\bf 43B} (1973) 245.

\bibitem{Salomon:1983xn} 
  M.~Salomon, D.~F.~Measday, J.~M.~Poutissou and B.~C.~Robertson,
  Nucl.\ Phys.\ A {\bf 414} (1984) 493.

\bibitem{Frlez:1997qu} 
  E.~Frle\v z {\it et al.},
  Phys.\ Rev.\ C {\bf 57} (1998) 3144
  [hep-ex/9712024].

\bibitem{Joram:1995gr} 
  C.~Joram {\it et al.},
  Phys.\ Rev.\ C {\bf 51} (1995) 2144.

\bibitem{Fitzgerald:1986fg}
  D.~H.~Fitzgerald {\it et al.},
  Phys.\ Rev.\ C {\bf 34} (1986) 619.

\bibitem{Mekterovic:2009kw} 
  D.~Mekterovi\'c {\it et al.} [Crystal Ball Collaboration],
  Phys.\ Rev.\ C {\bf 80} (2009) 055207
  [arXiv:0908.3845 [hep-ex]].

\bibitem{Jia:2008rt} 
  Y.~Jia, T.~P.~Gorringe, M.~D.~Hasinoff, M.~A.~Kovash, M.~Ojha, M.~M.~Pavan, S.~Tripathi and P.~A.~$\dot{\text{Z}}$o\l nierczuk,
  Phys.\ Rev.\ Lett.\  {\bf 101} (2008) 102301
  [arXiv:0804.1531 [nucl-ex]].

\bibitem{Blecher:1979zz} 
  M.~Blecher {\it et al.},
  Phys.\ Rev.\ C {\bf 20} (1979) 1884.

\bibitem{Bagheri:1987kc} 
  A.~Bagheri, K.~A.~Aniol, F.~Entezami, M.~D.~Hasinoff, D.~F.~Measday, J.~M.~Poutissou, M.~Salomon and B.~C.~Robertson,
  Phys.\ Rev.\ C {\bf 38} (1988) 885.

\bibitem{Auld:1979yd} 
  E.~G.~Auld {\it et al.},
  Can.\ J.\ Phys.\  {\bf 57} (1979) 73.

\bibitem{Moinester:1978zu} 
  M.~A.~Moinester {\it et al.},
  Phys.\ Rev.\ C {\bf 18} (1978) 2678.
 
\bibitem{Meissner:1997ii}
  U.-G.~Mei{\ss}ner and S.~Steininger,
  Phys.\ Lett.\ B {\bf 419} (1998) 403
  [hep-ph/9709453].
 
\bibitem{Fettes:1998wf}
  N.~Fettes, U.-G.~Mei{\ss}ner and S.~Steininger,
  Phys.\ Lett.\ B {\bf 451} (1999) 233
  [hep-ph/9811366].
  
\bibitem{Fettes:2000vm}
  N.~Fettes and U.-G.~Mei{\ss}ner,
  Phys.\ Rev.\ C {\bf 63} (2001) 045201
  [hep-ph/0008181].
  
\bibitem{Fettes:2001cr}
  N.~Fettes and U.-G.~Mei{\ss}ner,
  Nucl.\ Phys.\ A {\bf 693} (2001) 693
  [hep-ph/0101030].
  
\bibitem{Gridnev:2004mk}
  A.~B.~Gridnev, I.~Horn, W.~J.~Briscoe and I.~I.~Strakovsky,
  Phys.\ Atom.\ Nucl.\  {\bf 69} (2006) 1542
  [hep-ph/0408192].
  
\bibitem{Gibbs:1995dm}
  W.~R.~Gibbs, L.~Ai and W.~B.~Kaufmann,
  Phys.\ Rev.\ Lett.\  {\bf 74} (1995) 3740.
  
\bibitem{Matsinos:1997pb}
  E.~Matsinos,
  Phys.\ Rev.\ C {\bf 56} (1997) 3014.
  
\bibitem{Leutwyler:2015jga}
  H.~Leutwyler,
  PoS CD {\bf 15} (2015) 022
  [arXiv:1510.07511 [hep-ph]].

\end{thebibliography}
\end{document}